# The existence of a quantum phase transition in a Hubbard model on a quasi-one-dimentional two-leg ladder.

Valentin Voroshilov

Physics Department, Boston University, Boston, MA, 02215, USA

A canonical transformation of a new type is offered as the mean for studying properties of a system of strongly correlated electrons. As an example of the utility of the transformation, it is used to demonstrate the existence of a quantum phase transition in a Hubbard model on a quasi-one-dimensional two-leg ladder.

74.20.Mn    71.10.Li

## Motivation and the Hamiltonian

Since the first high temperature superconductor was discovered[1], there is no yet a commonly accepted explanation of this phenomenon. Many approaches are based on the Hubbard model[2]. The reason for using the Hibbard model is the fact that the parent state of a HTSC is an antiferromagnetic, which, when doped, exhibits many peculiar properties, including HTSC. The search for new non-perturbative methods[3] might lead to new insights on the matter and help to advance understanding of the nature of HTSC.

The first natural assumption one can make is that the magnetic order plays an important role in a HTSC compound. One might ask, however, if a magnetic order is important, why there is no HTSC based on a ferromagnetic parent material? In a ferromagnetic there is a strong average magnetization, which makes any small deviation from it being not significant. However, in an antiferromagnetic, any deviation from the average magnetization is not small.

Valentin Voroshilov, valbu@bu.edu, Physics Department, Boston University, 590 Commonwealth Ave., Boston, MA 02215

Hence, one can assume that having small deviations from the average magnetization is an important part of the phenomenon.

In an ideal antiferromagnetic, small deviations in magnetization are observed at temperatures slightly above the absolute zero and present themselves in a form of spin waves. Considering that temperature and doping affect the entropy of a system in a similar manner, one can assume, that in a doped antiferromagnetic at zero temperature the system also should exhibit spin waves.

Spin waves at the zero temperature in a doped antiferromagnetic could be used them as the source for effective electron attraction; however, the magnon theory of HTSC (or any other theory based on some mechanism leading to an effective electron attraction)[4], does not explain all the properties of HTSC.

Another possibility is to assume that spin waves govern all the important properties of HTSC and the superconductivity is just a "side effect" of the correlations between spin waves.

In an ideal antiferromagnetic two spin waves might travel in a correlated manner along two parallel chains of sites, because the presence of correlations decreases the entropy of the system. What happens if in a doped antiferromagnetic one wave encounters a defect in the form of a missing (or extra) electron? One possible outcome is breaking the correlation between two waves and increasing the entropy of the system. Another possibility is making an electron jump in order to preserve the presence of two sin waves traveling in a correlated manner. Such a jump would represent the interplay between magnetic and transport properties of a system. Obviously, one cannot rely on the above speculations to explain properties of an actual HTSC, but one could use these speculations to circumscribe the system for the further investigation. Trying to advance understanding of the interactions between strongly correlated electrons occupying two neighboring rows of sites of a lattice, one can start from a Hubbard model[5] for a 2-by-$N/2$ lattice (with N to be the total number of sites). Due to the assumption that strong interaction between neighboring electrons is responsible for "abnormal" properties of the system, at this point we will not include term describing on-site interaction (with an understanding that it is physically infeasible to suppress this kind of interaction).

Let us numerate the sites of a 2x($N/2$) lattice with $i = 1,...,(N/2)$ and $j = 1, 2,$ and using standard symbolism write the Hamiltonian:

$$H = -t\sum_{i,\sigma}(a^+_{i2\sigma}a_{i1\sigma} + HC) - t\sum_{i,j,\sigma}(a^+_{ij\sigma}a_{i+1j\sigma} + HC) + V\sum_{i,\sigma}n_{i1\sigma}n_{i2-\sigma}$$
$$+ V\sum_{i,j,\sigma}n_{ij\sigma}n_{i+1j-\sigma} + U\sum_{i,\sigma}n_{i1\sigma}n_{i2\sigma} + U\sum_{i,j,\sigma}n_{ij\sigma}n_{i+1j\sigma}.$$
(1)

To take into an account possible difference in the effective interaction between electrons with parallel spins and electrons anti-parallel spins we introduce a double-parametric potential energy term.

**The Canonical Transformation**

The challenge for understanding the properties of the system is to describe correlations happening in the real space and - at the same time - transport properties of the system related to the momentum space of the system.

In this paper we offer a canonical transformation of a new type, which is a composition of two canonical transformations: the first transformation happens in the real space and involves electrons from two neighboring sites lying on the opposite rows, and the second one makes a transition to the momentum space of the system.

The first transformation introduces a set of new Fermi operators $b_{ik}, b^+_{ik}$ ($k = 1,2,3,4$) (the new operators combine the creation and annihilation operators for electrons sitting "across" each other on two neighboring rows which lie on opposite sites):

$$a_{i1+} = -ub_{i1} + ub_{i2} - ub_{i3} - ub_{i4} + vb^+_{i1} + vb^+_{i2} - vb^+_{i3} + vb^+_{i4},$$

$$a_{i1-} = ub_{i1} - ub_{i2} - ub_{i3} - ub_{i4} + vb^+_{i1} + vb^+_{i2} + vb^+_{i3} - vb^+_{i4},$$
(2)

$$a_{i2+} = -ub_{i1} - ub_{i2} - ub_{i3} + ub_{i4} + vb^+_{i1} - vb^+_{i2} - vb^+_{i3} - vb^+_{i4},$$

$$a_{i2-} = -ub_{i1} - ub_{i2} + ub_{i3} - ub_{i4} - vb^+_{i1} + vb^+_{i2} - vb^+_{i3} - vb^+_{i4}.$$

Together with four more equations written for $a^+_{ij\sigma}$, and with two positive parameters $u, v > 0$ connected by condition $u^2 + v^2 = \frac{1}{4}$, equations (2) provide a canonical transformation (clearly, values $v = 0$ and $u = 0$ must be excluded from the set of the available values).

The second transformation is a standard transition from real space operators to operators acting in a momentum space (for this model calculation Planck's constant and a lattice constant are set to unity, and we assume periodic boundary conditions):

$$b_{lk} = \frac{1}{\sqrt{\frac{N}{2}}} \sum_p e^{ipl} b_{pk}, \qquad b_{lk}^+ = \frac{1}{\sqrt{\frac{N}{2}}} \sum_p e^{-ipl} b_{pk}^+ . \tag{3}$$

The values for momentum are confined by set $p = \frac{4\pi n}{N}, n = 0, \pm 1, ..., \pm \frac{N}{4}$.

To estimate the ground state energy of the system we use a well known variational approach[6]. First, we define a probe state vector of the ground state in the following form:

$$|E_0> = \prod_{|p|<\Pi} b_{p1}^+ b_{p2}^+ b_{p3}^+ b_{p4}^+ |0>$$ $(0 < \Pi < \pi)$, with $|0>$ to be the vacuum for operators $b_{pk}$, i.e. $b_{pk}|0> = 0$ (note: contrary to BCS[7] theory, we do not presume any paring in the system). Equation (2) does not conserve the number of particles in the system, hence, we demand that the expectation value of the operator for the total number of electrons is equal to the actual number of electrons in the system, $N_e$;

$$N_e = <E_0 | \sum_{ij\sigma} n_{ij\sigma} | E_0 > . \tag{4}$$

A routine calculation leads to the following system:

$$\frac{<E_0|H|E_0>}{16NU} = -\frac{T}{\pi}(\frac{1}{4} - 2y)\sin(\pi x) + 3x^2(\frac{1}{4} - y)^2 - 2(\frac{\sin(\pi x)}{\pi})^2(\frac{1}{4} - y)^2 +$$
$$+ 3(1-x)^2 y^2 - 2(\frac{\sin(\pi x)}{\pi})^2 y^2 + 6y(\frac{1}{4} - y)x(1-x) + 4y(\frac{1}{4} - y)(\frac{\sin(\pi x)}{\pi})^2 + \tag{5}$$
$$+ W\left\{3x^2(\frac{1}{4} - y)^2 + 3(1-x)^2 y^2 + y(\frac{1}{4} - y)(x^2 + (1-x)^2) + 4y(\frac{1}{4} - y)x(1-x)\right\},$$

$$x = \frac{\frac{1}{8}\frac{N_e}{N} - y}{\frac{1}{4} - 2y}, \tag{6}$$

with the following definitions and conditions

$$W = \frac{V}{U}, \quad 0 < x = \frac{\Pi}{\pi} < 1, \quad 0 < T = \frac{t}{U}, \quad 0 < y = v^2 < \frac{1}{4}, \quad 0 < \frac{N_e}{N} < 2. \quad (7)$$

When $u, v \neq 0$ an anomalous electron pair correlation function for electrons with anti-parallels spins is not zero; $< E_0 | a_{i1+} a_{i2-} | E_0 > = 4uv(1-2x)$. This is a clear sign of a new phase in the system. This new phase can be reached only when parameters of the Hamiltonian satisfy certain conditions and lead to the minimum of the ground state energy.

Analysis shows that when $W < 0$, Eq. (5) constrained by conditions (6) and (7) does not have a global minimum, which means transformation (2) is not applicable for this case. Within this model an effective electron attraction between electrons with anti-parallel spins destroys the possibility for a new phase.

When $W = 1$ there is a very narrow set of values for $T$ and $\frac{N_e}{N}$ for which transformation (2) can be applied (when $T < 0.3$ the global minimum is only reached at $y = 0$, but increasing $T$ leads to a vary narrow set of available values for $\frac{N_e}{N}$, even when $T = 1$ transformation (2) cannot be applied if $0.09 < \frac{N_e}{N} < 1.92$). However, when $W < 1$, the region of values allowing the new phase widens (for example, if $W = 0.5$ and $T = 0.8$ transformation (2) applies when $0 < \frac{N_e}{N} < 0.14$ or $1.87 < \frac{N_e}{N} < 2$). This could be seen as an indicator of the influence magnetic properties of the system (i.e. spin-spin interaction which affect values $U$ and $V$) exert on its transport properties. The widest region of the values allowing transformation (2) is achieved when $W = 0$ (electrons with antiparallel spins are not present in the Hamiltonian), which indicates that correlations between electrons with parallel spins might be primarily responsible for the existence of the new phase (despite the fact that $< E_0 | a_{i1+} a_{i2+} | E_0 > = 0$).

In conclusion, it is shown that for a certain range of the parameters of Hamiltonian (1), the model shows properties which resemble some of the properties of HTSC, namely, the existence of anomalous electron pair correlation functions. It is worth to underline there has not been hypothesizing about a paring mechanism and the probe ground state is not based on an assumption for an existence of correlated electron pairs, which supports the view that the mechanism of HTSC is drastically different from the classical BCS[8] mechanism.

Obviously, using the canonical transformation described above one cannot yet make any conclusions on the physical properties of actual systems. The model must include on-site interaction and be generalized to a two-dimensional situation; the analysis has to be extended above the ground state and provide information about an excitation spectrum. A standalone mathematical task is analyzing properties of transformation (2) and its generalizations, which is the goal of an ongoing investigation.